\documentstyle[12pt,epsf,cite]{article}
\begin{document}

\title{Volume dependence of the phase boundary in 4D dynamical
  triangulation} \author{Bas V. de Bakker\thanks{email:
    bas@phys.uva.nl} \\Jan Smit\thanks{email: jsmit@phys.uva.nl} \and
  Institute for Theoretical Physics, University of Amsterdam\\ 
  Valckenierstraat 65, 1018 XE Amsterdam, the Netherlands.}

\date{13 May 1994}

\maketitle

\begin{abstract} 
  The number of configurations of the dynamical triangulation model of
  4D euclidean quantum gravity appears to grow faster than
  exponentially with the volume, with the implication that the system
  would end up in the crumpled phase for any fixed $\kappa_2$ (inverse
  bare Newton constant).  However, a scaling region is not excluded if
  we allow $\kappa_2$ to go to infinity together with the volume.
\end{abstract}

{
\thispagestyle{myheadings}
\renewcommand{\thepage}{ITFA-94-14}

\clearpage
}

\section{Introduction}

The dynamical triangulation model of four dimensional euclidean
quantum gravity has recently been the subject of  
interest \cite{migdal1,migdal2,aj,Var93,Brue,kogut1,kogut2}.

The  canonical partition function of the model at some fixed 
volume (number of four-simplices) $V$ may be defined as a 
sum over triangulations  ${\cal T}$ of the hypersphere $S^4$  
\begin{equation}
  Z(V,\kappa_2) = \sum_{{\cal T}(N_4 = V)} \exp(\kappa_2 N_2),
  \label{partv}
\end{equation}
where $N_i$ is the number of simplices of dimension $i$ in the
triangulation ${\cal T}$ with fixed edge lengths.  The $N_i$
$(i=0\ldots4)$ satisfy three constraints which means that only two of
them are independent.  We have chosen $N_2$ and $N_4$ as the
independent variables. For comparison with other work we remark that
if $N_0$ is chosen instead of $N_2$ then the corresponding coupling
constant $\kappa_0$ is related to our $\kappa_2$ as $\kappa_0 = 2
\kappa_2$.

The weight $\exp(\kappa_2 N_2)$ is part of the Regge-Einstein action
\begin{eqnarray}
  S & = & \frac{-1}{16\pi G_0} \sum_{\Delta} V_{2}\, R_{\Delta} =
  \kappa_2 (\rho N_4 - N_2),\label{RE}\\ \kappa_2 &=& \frac{V_2}{8
    G_0},\;\;\; \rho= \frac{10\arccos(1/4)}{2\pi},
\end{eqnarray}
where $V_2=(324/5)^{-1/4}$ is the volume of 2-simplices (triangles
$\Delta$) in units of $V_4=1$.  Numerical simulations have shown that
the system at fixed volume can be in two phases. For $\kappa_2 >
\kappa_2^c(N_4)$ (weak bare coupling $G_0$) the system is in an
elongated phase with high $\langle R_{\Delta}\rangle$ while for
$\kappa_2 < \kappa_2^c(N_4)$ (strong coupling) it is in a crumpled
phase with low $\langle R_{\Delta}\rangle$. The elongated phase has a
low effective dimensionality and a large average distance while the
crumpled phase has a high effective dimensionality and a small average
distance between the simplices.  The transition at $\kappa_2^c$ is
found to be continuous with a susceptibility $\partial^2 \ln
Z(N_4,\kappa_2)/{\partial \kappa_2}^2$ that grows with $V$. This opens
the exciting possibility that continuum behavior may be found at the
phase boundary \cite{migdal1,migdal2,aj,Var93,Brue,kogut1}.

Recently, evidence was presented \cite{kogut2} that the partition 
function $Z(N_4,\kappa_2)$ grows faster than exponentially in $N_4$ 
at fixed $\kappa_2$. This result was interpreted as to cast doubt on 
the dynamical triangulation approach to four dimensional gravity, 
because it implies that the grand canonical partition function
\begin{equation}
  Z(\kappa_2,\kappa_4) = \sum_{{\cal T}} \exp (\kappa_2 N_2 - \kappa_4 N_4)
\end{equation}
is ill defined (the parameter $\kappa_4$ is related to the bare 
cosmological constant). We like to argue however that it is 
useful to study the local and ultraviolet properties of the model 
separately from the global and cosmological properties. The 
nonexistence of $Z(\kappa_2,\kappa_4)$ need not invalidate the 
possibility of a scaling regime at large $N_4$. In this letter we 
report on our results for the behavior of $Z(N_4,\kappa_2)$ and 
shall argue that a possible continuum limit involves sending also 
$\kappa_2$ to infinity, i.e. $G_0 \to 0$. 

\section{Method}

Unfortunately, no set of local moves to simulate the canonical
ensemble is known and probably no such set exists \cite{nabutovsky}.
Therefore, we rewrite the partition function (\ref{partv}) as
\begin{equation}
  Z(V,\kappa_2) = \exp(\Delta S(V)) \sum_{\cal T} \exp(\kappa_2 N_2 -
  \Delta S(N_4)) \delta_{N_4,V},
  \label{partfunc}
\end{equation}
where the sum is now over all triangulations ${\cal T}$  with 
$S^4$ topology.  In a
simulation we can then generate configurations with Boltzmann weight
\begin{equation}
  \exp(\kappa_2 N_2 - \Delta S(N_4))
  \label{weights}
\end{equation}
and select only those configurations with $N_4=V$.
The precise form of $\Delta S$ is unimportant.  We have taken
followed \cite{migdal1,migdal2} and used
\begin{equation}
  \Delta S(N_4) = -\kappa_4 N_4 - \gamma (N_4-V)^2,
  \label{deltas}
\end{equation}
with $\gamma = 5 \cdot 10^{-4}$.  This is smaller than the value used
in \cite{kogut1,kogut2}, resulting in somewhat larger volume
fluctuations.  As the number of triangulations cannot grow faster than
$(5N_4)! \sim \exp(5 N_4 \ln N_4)$ such a term
ensures that the volume does not blow up in the simulation.
The interesting question is now how $Z(V,\kappa_2)$ behaves as a
function of the volume and the coupling constant $\kappa_2$.

The simulation of the system 
\begin{eqnarray}
  Z' & = & \sum_{\cal T} \exp(\kappa_2 N_2 - \kappa_4 N_4 - \gamma
  (N_4 - V)^2) \\
  & = & \sum_{N_4} \exp(\ln Z(N_4,\kappa_2) - \kappa_4 N_4 - \gamma (N_4 - V)^2)
\end{eqnarray}
allows the measurement of $\partial \ln Z/\partial N_4$ through the
equation
\begin{equation}
  \kappa_4^c(V,\kappa_2) \equiv \frac{\partial \ln
    Z(V,\kappa_2)}{\partial V} = \kappa_4 + 2 \gamma (\langle N_4
  \rangle - V) + O\left(\frac{1}{V}\right),
\end{equation}
which follows from a saddle point approximation for
$\langle N_4
\rangle$ in this ensemble.  At the same time this simulation gives us
configurations with the weights (\ref{weights}) to measure properties
of (\ref{partfunc}). 

We will not expand here upon our method to generate simplicial
complexes, but for definiteness let us mention that we allow all those
complexes where no two $d$-dimensional (sub)simplices share exactly
the same set of $(d+1)$ points. (In \cite{BaSm94} our criteria were
somewhat different, our present criteria lead to numerical results in
agreement with \cite{migdal2,kogut1,kogut2,aj,Brue}.)

\section{Results}

\begin{figure}[t]
  \epsfxsize=\textwidth
  \epsffile{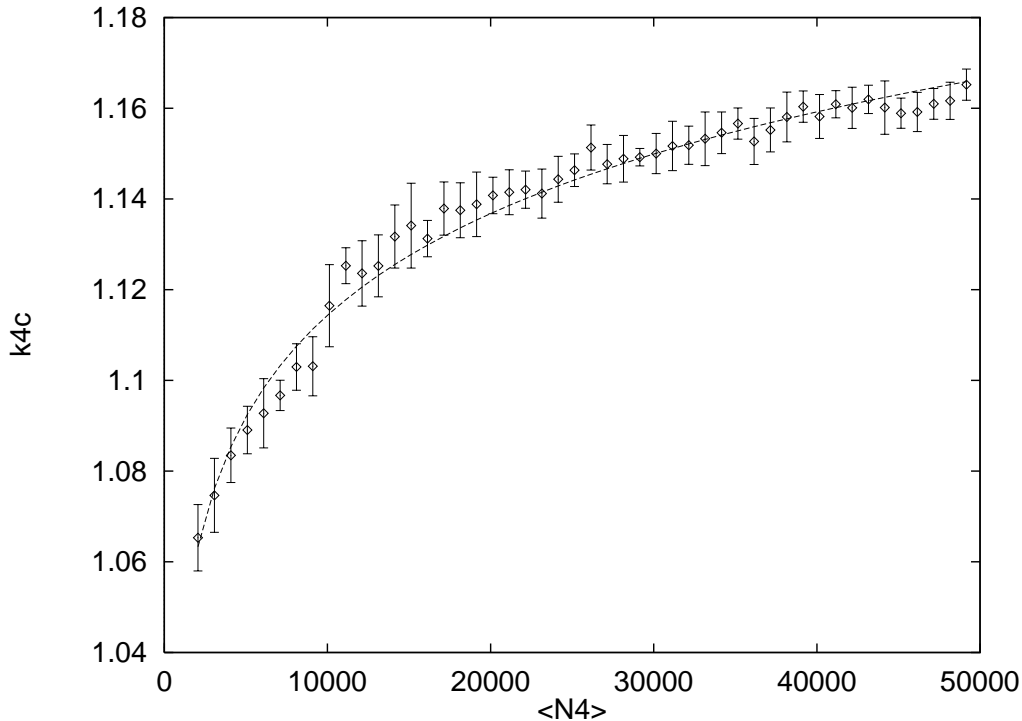}
  \caption{The effective critical $\kappa_4^c$ for various 
    volumes at $\kappa_2 = 0$.}
\label{andiv}
\end{figure}

Figure~\ref{andiv} shows the value of $\kappa_4^c$ as a function of
the volume at $\kappa_2 = 0$, where the configurations at any
particular volume all have the same weight.  The points at the highest
volumes are somewhat correlated as the autocorrelation time became of
the order of the time between volume increments. The line is a fit to
\begin{equation}
  \kappa_4^c = a + b \ln \langle N_4 \rangle,
  \label{logfit}
\end{equation}
where the fitting parameters are 
\begin{equation}
  a  =  0.82(1),\;\;\; b  =  0.0323(9), \label{fitb} 
\end{equation}
with $\chi^2 = \mbox{24 at 46 d.o.f.}$
We have also tried fitting the data to the converging functions
\begin{equation}
  \kappa_4^c = \sum_{n=0}^{k} a_n \langle N_4 \rangle ^{-n}
  \label{sumpowfit}
\end{equation}
for various small $k$ and
\begin{equation}
  \kappa_4^c = a + b \langle N_4 \rangle ^{-c}.
  \label{smallpowfit}
\end{equation}
The function (\ref{sumpowfit}) simply does not yield any reasonable
fit.  On the other hand, the small converging power
(\ref{smallpowfit}) fits with $a = 1.7(9)$ and $c = 0.06(8)$, which is
consistent with zero.  This also points to a logarithm.  The large
error in $a$ is due to its large correlation with the power $c$.  For
fixed $c$ the error in $a$ would be much smaller.  Both the
qualitative picture and the value of $b$ (\ref{fitb}) agree with data
reported in \cite{kogut2}, produced via completely independent code.

The forms (\ref{logfit}) and (\ref{smallpowfit}) are virtually
indistinguishable for small powers.  Thus, if it is indeed a
logarithm, a converging small power can never be excluded and will
even be favored if the low volume points have somewhat lower
$\kappa_4^c$ than that of a pure logarithm.

At the other side of the transition, at $\kappa_2 = 2.0$, the same
plot (not shown) is just a horizontal line for all volumes we have
used, which went up to 90,000.  The fitting parameters for the
logarithmic function (\ref{logfit}) are  in this case
\begin{equation}
  a  =  5.662(5),\;\;\; b  =  1(5) \times 10^{-4},\label{fitb2}
\end{equation}
with $\chi^2 = \mbox{7.7 at 86 degrees of freedom}$, which is
consistent with a constant (i.e.\ $b$ is consistent with 0).  So in
this phase the number of configurations making an important
contribution to the partition function does rise only exponentially
with the volume.

\begin{figure}[t]
  \epsfxsize=\textwidth
  \epsffile{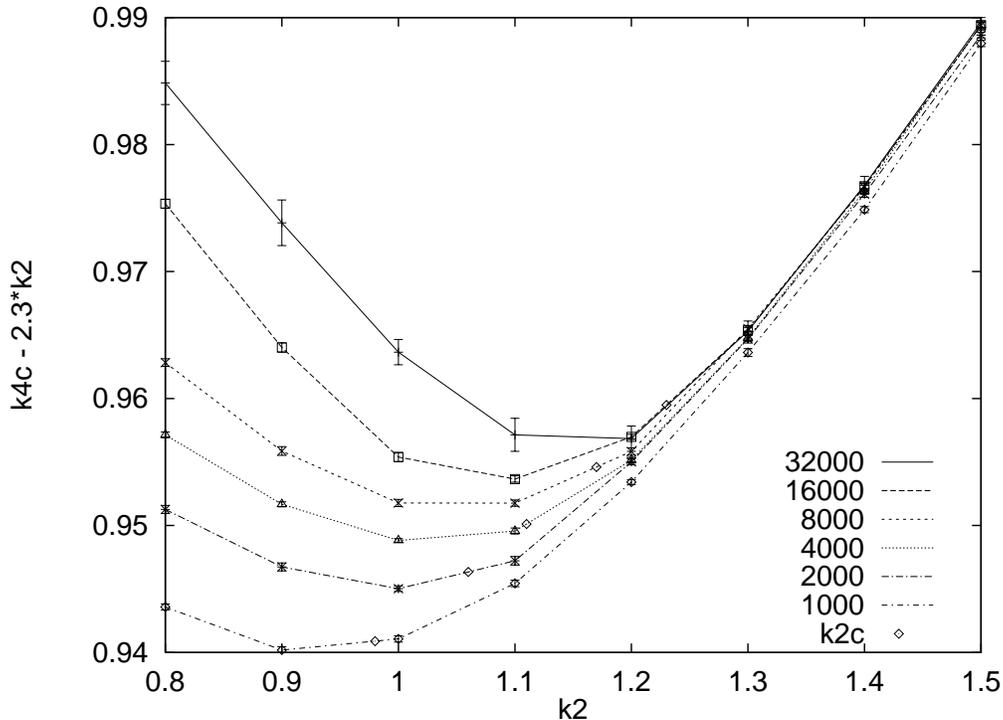}
  \caption{The  effective critical $\kappa_4^c$ as a function of 
    $\kappa_2$ at various volumes.  To expand the vertical scale we
    have subtracted a straight line $-2.3 \kappa_2$ from the data. The
    values of $\kappa_2^c(V)$ are indicated with diamonds.}
  \label{k4ck2}
\end{figure}

The values of $\kappa_4^c$ at various volumes and $\kappa_2$ values in
the region of the transition can be seen in figure~\ref{k4ck2}.  We
have subtracted $2.3\kappa_2$ from the data for $\kappa_4^c$ to expand
the vertical scale.  The diamonds indicate the value of $\kappa_2^c$
at each volume.  For the lowest three volumes these are the points
obtained in \cite{kogut1}, while those of 8000 and 16000 are our own.

Going from right to left the curves are close together and then break
away from each other as their slope decreases.
The break away point moves to the right as the the volume increases.
Assuming that it is correct to extrapolate the pattern in this figure
and eqs. (\ref{fitb},\ref{fitb2}) to large volumes, we conclude that
$\kappa_4^c$ first stays constant as the volume increases and then,
depending on the value of $\kappa_2$, starts to
diverge logarithmically.
We also see that the transition between the crumpled and 
elongated phase moves to larger $\kappa_2$ as the volume 
increases, staying near the break away points.

There is a close relation between these curves and the average
curvature, due to the relation
\begin{equation}
  \frac{\partial \langle N_2 \rangle}{\partial V} = \frac{\partial^2
    \ln Z(V, \kappa_2)}{\partial \kappa_2 \partial V} = \frac{\partial
    \kappa_4^c}{\partial \kappa_2}.
\end{equation}
This implies that the largest change in the slope of these curves will
coincide with the susceptibility peak and that as that peak becomes
narrower the bend in the $\kappa_4^c(\kappa_2)$ curve becomes sharper.

\section{Discussion}

These results indicate that the number of crumpled configurations
grows factorially with the volume, while the number of elongated
configurations only grows exponentially.  This would mean that at any
fixed $\kappa_2$ the crumpled configurations will always dominate for
large enough volumes.

From this it already follows that the $\kappa_2^c$ of the transition
between the phases must diverge with the volume.  In ref.
\cite{kogut1} a converging scenario is favored but a diverging one not
excluded.  In fact, it is mentioned \cite{kogut1} that leaving out
data points with $V=500,1000$ leads to fit with $\kappa_2^c$ diverging
logarithmically with volume. The large $\kappa_2$ region in fig. 2
seems indeed to suggest that $V=1000$ is somewhat too low to see the
asymptotic trend.

It is an interesting question whether the bare free energy
\begin{eqnarray}
  F(V,\kappa_2) = - \ln \sum_{{\cal T} (N_4 = V)} \exp (-S)
\end{eqnarray}
is extensive at the phase boundary, $\partial
F(V,\kappa_2^c(V))/\partial V = const\,$.  From (\ref{RE}) we see that
this would be the case if
\begin{equation}
  \kappa_4^c(V,\kappa_2^c(V))- \rho \kappa_2^c(V)-\rho V
  \frac{\partial \kappa_2^c(V)}{\partial V} + \langle N_2 \rangle
  \frac{\partial \kappa_2^c(V)}{\partial V}
\end{equation}
is independent of $V$ for large $V$.  Assuming a logarithmic
dependence of $\kappa_2^c$ on $V$ the last two terms are constant so
the question is whether the difference of the first two is $V$
independent. We see from fig. 2 that this unlikely: in the range $V =
2000 - 16000$, $\kappa_4^c \approx 2.4 \, \kappa_2^c(V)$ which rises
faster with than $\rho\kappa_2 \approx 2.098 \, \kappa_2$.

Expecting extensivity with the bare Einstein-Regge action is 
questionable. It consists of two very different terms $\propto 
N_2$ and $\propto N_4$ which will require different 
renormalization, such that the particular linear combination 
$\kappa_2 (N_2- \rho N_4)$ has to be modified. In other words, the 
curvature term will mix with the volume $N_4$ under 
renormalization. A natural candidate is $\kappa_2 N_2 - 
\kappa_4^c(N_4,\kappa_2) N_4 + const\, N_4$, where the $const$  
may be fixed by some normalization condition.

We are therefore not disturbed by the fact that the average bare
curvature is rather different from zero at the transition \cite{aj}.
One may also contemplate obtaining a physical curvature in terms of a
physically defined metric, but there are other perhaps more easily
accessible quantities. Elsewhere \cite{BaSm94} we have proposed a way
of measuring the renormalized Newton constant $G$ in terms of the
binding energy of test particles. It is however not even clear to us
that $V$ should be interpreted as the physical volume as measured in
terms of a physical metric at physical scales, because $V$ is
sensitive to the proliferation of baby universes \cite{ajjk}.

We have presented results suggesting that the bare gravitational
coupling $G_0$ has to go to zero in a possible scaling limit.  It is
interesting that also in matrix models of 2D gravity with unrestricted
topology the number of configurations rises factorially with the
volume and a sensible continuum limit can only be taken by letting the
bare $G_0$ go to zero in the so called double scaling limit
\cite{DS,GM,BK}.  Perhaps 2D gravity in its dynamical triangulation
formulation (without restriction on topology) has also a strong
coupling phase.

Similarly, the scenario with $G_0$ going to zero might also be
applicable to the case of four-dimensional simplicial gravity with
unrestricted topology.  In this model it is certain that there is no
exponential bound on the number of configurations.

\section*{Note}

Just as we finished this paper we received a preprint \cite{ajexp}
presenting data that favors the scenario with $\kappa_4^c$ converging
according to (\ref{smallpowfit}) with an exponent $c$ of $1/4$.


\begin{thebibliography}{99}

\bibitem{migdal1} M.E.~Agishtein and A.A.~Migdal, Mod.\ Phys.\ Lett.\ 
  A7 (1992) 1039.

\bibitem{migdal2} M.E.~Agishtein and A.A.~Migdal, Nucl.\ Phys.\ B385
  (1992) 395.

\bibitem{aj} J.~Ambj\o rn and J.~Jurkiewicz, Phys.\ Lett.\ B278 (1992)
  42.

\bibitem{Var93} S.~Varsted, Nucl.\ Phys.\ B412 (1994) 406.

\bibitem{Brue} B.~Br\"ugmann, Phys.\ Rev.\ D47 (1993) 3330.

\bibitem{kogut1} S.~Catterall, J.~Kogut and R.~Renken {\it Phase
    Structure of Four Dimensional Simplicial Quantum Gravity}, CERN
  preprint CERN-TH.7149/94 and Illinois preprint ILL-(TH)-94-26.

\bibitem{kogut2} S.~Catterall, J.~Kogut and R.~Renken {\it On the
    absence of an exponential bound in four dimensional simplicial
    gravity}, CERN preprint CERN-TH.7197/94 and Illinois preprint
  ILL-(TH)-94-07.

\bibitem{nabutovsky} A.~Nabutovsky and R.~Ben-Av, Comm.\ Math.\ Phys.\
  157 (1993) 93.

\bibitem{BaSm94} B.V.~de Bakker and J.~Smit, Nucl.\ Phys.\ B (Proc.\ Suppl.)
  34 (1994) 739.

\bibitem{ajjk} J.~Ambj\o rn, S.~Jain, J.~Jurkiewicz and
  C.F.~Kristjansen, Phys.\ Lett.\ B305 (1993) 208.

\bibitem{DS} M.~Douglas and S.~Shenker, Nucl.\ Phys.\ B335 (1990) 635.

\bibitem{GM} D.J.~Gross and A.A.~Migdal, Nucl.\ Phys.\ B340 (1990) 333.

\bibitem{BK} E.~Br\'ezin and V.A.~Kazakov, Phys.\ Lett.\ B236 (1990)
  144.

\bibitem{ajexp} J.~Ambj\o rn and J.~Jurkiewicz {\it On the exponential
    bound in four dimensional simplicial gravity}, preprint
  NBI-HE-94-29.

\end{thebibliography}
\end{document}